# Chaotic Dynamic S-Boxes Based Substitution Approach for Digital Images

**Musheer Ahmad and Akshay Chopra**
Department of Computer Engineering, Faculty of Engineering and Technology,
Jamia Millia Islamia, New Delhi 110025, India

*ABSTRACT*

*In this paper, we propose an image encryption algorithm based on the chaos, substitution boxes, nonlinear transformation in galois field $GF(2^8)$ and Latin square. Initially, the dynamic S-boxes are generated using Fisher-Yates shuffle method and piece-wise linear chaotic map. The algorithm utilizes advantages of keyed Latin square and affine-power-affine transformation to substitute highly correlated digital images and yield encrypted image with valued performance. The chaotic behaviour is achieved using Logistic map which is used to select one of thousand S-boxes and also decides the row and column of selected S-box. The selected S-box value is transformed using nonlinear affine-power-affine transformation. Along with the keyed Latin square generated using a 256-bit external key, used to substitute secretly plain image pixels in cipher block chaining mode. To further strengthen the security of algorithm, round operation are applied to obtain final ciphered image. The experimental results are performed to evaluate algorithm and the anticipated algorithm is compared with a recent encryption scheme. The analyses demonstrate algorithm's effectiveness in providing high security to digital media.*

## 1. INTRODUCTION

With the advent and continuous development in the field of network communication, the need for secure transfer of multimedia data over the Internet has been increasing at a very high rate. Internet provides a medium of exchange as well as communication for people around the world. But, the openness of Internet poses various threats to digital data transmitted through public Internet. The openness of public networks calls for the need of certain mechanism that can ensure end-to-end secrecy of digital media. To serve the purpose, the cryptographic systems are developed for secret storage, writing and transmission. It encompasses various aspects of information security such as confidentiality, integrity, non-repudiation and authentication thus ensuring the privacy of information during storage and transmission.

Substitution boxes are indispensable nonlinear component of modern day cryptographic systems, widely used in symmetric-key encryption algorithms like DES, IDEA, AES, Blowfish, KASUMI, RC5, Lucifer, GOST, etc. The vital role of S-box is to abstruse the relationship between ciphertext and the secret key, which is also called the Shannon's property of confusion. S-boxes are primarily meant to introduce nonlinearity in encryption algorithms, thereby making them resistant to linear and differential cryptanalysis [1]. Thus S-box forms the core part of encryption algorithms. The strength of these algorithms is entirely dependent on the amount of confusion introduced by the S-box. The performance of the S-box is highly dependent on the area of use and also on the nature of data. Since Rinjdael S-box assumes to have excellent features and plays a critical role in the success of AES. Many researchers have focused their research on designing and improving the features, like nonlinearity, of S-box [2]. In [3] Cui and Cao designed a new S-box structure named affine-power-affine to improve the algebraic complexity of original Rinjdael AES S-box and making it more stronger as compared to Rinjdael S-box. But, when it comes to high auto-correlated data, as in the case of digital media like images, S-boxes show poor performance despite of having high nonlinearity [4]. This is because of the existence of extreme correlation among neighbouring pixels in images which leads to weak substitutions through S-box and may even depicts the correlation in substituted image as well. Thus attacking such substituted image becomes easy from cryptanalytic point of view.

In this work, we formulate a new image encryption algorithm which amalgamates the characteristics of chaotic systems, dynamic S-boxes generated using Fisher-Yates shuffle, keyed Latin square and affine-power-affine transformation. In order to overcome the problem of substitution of correlated image pixels, we use chaos to select randomly a dynamic S-box from 1000 generated S-boxes for pixels substitution. The proposed algorithm exhibits several advantages which includes high visual indistinguishability, high entropy, large key space, high plaintext sensitivity and low auto-correlation.

The rest of the paper is organized as follows: Section 2 gives an overview of the work done in field of S-box based image encryption along with their shortcomings. Section 3 gives a briefly describes the preliminaries used in the design of proposed algorithm. The proposed substitution based encryption algorithm along with the method of generating dynamic S-boxes is discussed in Section 4. In Section 5, the strength and performance of algorithm against standard tests is quantified and the results are analyzed and compared with a recent S-box substitution based encryption reported by Anees *et al.* to demonstrate the effectiveness of our algorithm. Finally, the conclusions of work are made in Section 6.

## 2. LITERATURE REVIEW

Wang *et al.* [12] proposed dynamic S-boxes based image encryption method which encrypts image by dividing it into blocks. A random S-box is generated per block which substituted all pixels of the block. The substitution process is performed from four sides in order to strengthen the algorithm. Although, this algorithm proved to be resistant to plaintext attacks, it is still not able to break the correlation that exists between the pixels of a particular group since the same S-box was used for all the pixels in a block. Zhang *et al.* [13] made an attempt to overcome the disadvantage of Wang *et al.* scheme by using different S-box for each plain image pixel to break completely the correlation of pixels in image. They proposed an encryption scheme based on chaos and alternate circular S-boxes constructed using a chaotic system. Where, for each plain image pixel, an S-box is selected according to plain image pixel and substitution is performed to get encrypted pixel. The substitution is performed in both forward as well as backward direction. After the each substitution, the S-box is randomly rotated cyclically and forming a new S-box. Eventually, the plain image pixels are get substituted by different S-boxes and encrypted image whose pixels are completely decorrelated is obtained.

## 3. PRELIMINARIES

### 3.1 Chaotic Logistic map

The 1D Logistic map proposed by May [14] in 1967 is one of the simplest nonlinear chaotic discrete systems that exhibits chaotic behavior; its state equation is governed by the following iterative formula.

$$x(n+1) = \lambda \times x(n) \times (1 - x(n)) \qquad (1)$$

Where $x$ is map's variable and $x(0)$ acts as its initial condition, $\lambda$ is system parameter and $n$ is number of iterations needs to be applied. Here $\lambda \in [0, 4]$. The research shows that when $\lambda \in [0, 3]$, the value of $x$ reaches a fixed value after several iterations without any showing any chaotic dynamics. When $\lambda \in (3, 3.57]$ the map oscillates between 2 fixed values without any chaotic dynamics. The map exhibits chaotic dynamics for $3.57 < \lambda < 4$ and $x(n) \in (0, 1)$ for all $n$. The initial value $\lambda$ and $x(0)$ acts as secret key when employed in an encryption system. To successfully decrypt the message, exact values of both $\lambda$ and $x(0)$ are needed at the receiver end. Thus, the algorithm becomes entirely key dependent which makes the extraction of information from the encrypted image difficult for the attacker.

### 3.2 Piece-wise Linear Chaotic Map

A 1D linear piecewise linear chaotic map composed of various multiple linear segments exhibiting good dynamical properties. This map is one of the most studied chaotic maps and is given by the following state equation [15].

$$y(n+1) = \begin{cases} \dfrac{y(n)}{p} & \text{if} \quad 0 < y(n) \le p \\ \dfrac{1-y(n)}{1-p} & \text{if} \quad p < y(n) < 1 \end{cases} \quad (2)$$

Where $x$ refers to the state of the system, $n$ refers to the number of iterations and $y(n) \in (0, 1)$. The system trajectory of PWLCM $y(n)$ visits the entire interval $(0, 1)$ for every value of parameter $p \in (0, 1)$ [16]. It has been studied that PWLCM exhibits good statistical features and it has no periodic windows (non-chaotic regions) [17]. Again, the initial values assigned to $y(0)$ and $p$, for execution of map, act as key to control its behaviour and trajectory. The piece-wise linear chaotic map is utilized to generate dynamic S-boxes.

### 3.3 Affine-Power-Affine Transformation

The Rijndael S-box referred to as the AES S-box was incorporated in the AES algorithm to bring non-linearity into the encryption system. The S-box was designed in such a way so as to make the whole system resistant to linear and differential cryptanalyses. It was done by diminishing correlation among the input as well as the output bits. Moreover, the concept of affine transform was added to protect it against algebraic cryptanalysis. It has been highlighted that the algebraic structure of AES S-box contains only 9 terms along with the reason, which makes it vulnerable to algebraic attacks [18]. Therefore, several attempts have been made to improve the algebraic complexity of the AES S-box to make algebraic attacks as difficult as possible. Cui and Cao, in 2007, designed a new structure referred to as the Affine-Power-Affine to aggrandize the algebraic complexity [3]. Consequently, the algebraic complexity gets bettered and the number of terms in the improved algebraic equation augmented from 9 to 253. The power function $P(x)$ (multiplicative inverse modulo $x^8 + x^4 + x^3 + x + 1$) and the affine-transformation $A(x)$ [1, 3] are defined as:

$$P(x) = \begin{cases} x^{-1} & x \ne 0 \\ 0 & x = 0 \end{cases}$$

$$A(x) = \begin{bmatrix} 1 & 0 & 0 & 0 & 1 & 1 & 1 & 1 \\ 1 & 1 & 0 & 0 & 0 & 1 & 1 & 1 \\ 1 & 1 & 1 & 0 & 0 & 0 & 1 & 1 \\ 1 & 1 & 1 & 1 & 0 & 0 & 0 & 1 \\ 1 & 1 & 1 & 1 & 1 & 0 & 0 & 0 \\ 0 & 1 & 1 & 1 & 1 & 1 & 0 & 0 \\ 0 & 0 & 1 & 1 & 1 & 1 & 1 & 0 \\ 0 & 0 & 0 & 1 & 1 & 1 & 1 & 1 \end{bmatrix} \times \begin{bmatrix} x_0 \\ x_1 \\ x_2 \\ x_3 \\ x_4 \\ x_5 \\ x_6 \\ x_7 \end{bmatrix} \oplus \begin{bmatrix} 1 \\ 1 \\ 0 \\ 0 \\ 0 \\ 1 \\ 1 \\ 0 \end{bmatrix}$$

Where $x_i$'s are the coefficients of $x$ (8-bit elements of S-box in $GF(2^8)$). The algebraic expression representing AES S-box is:

$$y = 05x^{FE} + 09x^{FD} + F9x^{FB} + 25x^{F7} + F4x^{EF} + 01x^{DF} + B5x^{BF} + 8Fx^{7F} + 63$$

The affine-power-affine transformation of an element $x$ of AES S-box is defined as [3]:

$$S(x) = A \circ P \circ A$$

The APA S-box obtained after applying the affine-power-affine transformation is shown in the table:

|   | 0  | 1  | 2  | 3  | 4  | 5  | 6  | 7  | 8  | 9  | A  | B  | C  | D  | E  | F  |
|---|----|----|----|----|----|----|----|----|----|----|----|----|----|----|----|----|
| 0 | 8C | 90 | D9 | C1 | 46 | 63 | 53 | F1 | 61 | 32 | 15 | 3E | 26 | 9A | 97 | 2E |
| 1 | D8 | 80 | 99 | 9E | C0 | 95 | 67 | B7 | 6D | E0 | F3 | 28 | 20 | 86 | B6 | EF |
| 2 | 4B | 31 | B5 | D2 | 13 | 39 | 6C | AF | 03 | 3F | 4D | 34 | F9 | EC | 8E | 17 |
| **3** | C5 | 25 | 3C | 89 | C9 | 2B | 3A | C2 | 6E | C6 | AA | 91 | 49 | 18 | 93 | DE |
| **4** | 0D | 6F | 65 | AF | 92 | A7 | F6 | A6 | 40 | B9 | ED | B0 | C3 | D7 | 7D | 7C |
| **5** | 54 | 59 | DF | 2F | DA | A4 | 05 | 94 | 9B | 72 | 01 | 74 | A9 | F7 | 81 | E9 |
| **6** | 1F | B3 | EB | CF | E8 | 47 | 52 | 36 | BC | 16 | 29 | 76 | 12 | FA | 9C | 8A |
| **7** | 5B | A8 | 43 | D1 | 79 | 85 | 42 | 82 | C7 | A1 | 78 | 4F | E2 | 35 | EA | AD |
| **8** | DC | 0E | D3 | 2D | 6A | 5A | 44 | AB | C8 | E5 | 37 | 0A | 6B | 51 | E3 | 14 |
| **9** | CD | 56 | 4A | D6 | 08 | 83 | BB | 33 | E1 | 30 | 4E | 24 | 5E | B4 | 00 | 48 |
| **A** | 5F | 22 | 0B | 50 | 3D | 80 | 1A | BF | CC | FF | 64 | 87 | 1B | C4 | 07 | F8 |
| **B** | 0C | D4 | AC | 02 | 10 | 84 | 7E | 69 | 70 | 60 | 55 | 2A | 21 | 57 | 23 | 66 |
| **C** | 62 | 73 | CB | 41 | 58 | 71 | 77 | 1C | 7B | 8F | 9F | 9D | A3 | B1 | 7F | 5D |
| **D** | F4 | 06 | AE | D5 | E6 | 3B | BA | FE | 96 | E7 | 0F | 45 | 2C | F0 | FC | BD |
| **E** | E4 | 98 | FB | CA | 11 | F5 | DD | 7A | 5C | FD | CE | 88 | D0 | 68 | 8D | 4C |
| **F** | BE | 04 | 38 | 1D | 1E | F2 | 27 | 19 | B2 | 75 | A2 | EE | DB | B8 | 09 | 8B |

Cui and Cao has shown that compared to AES S-box, the affine-power-affine S-box has stronger algebraic complexity. The coefficients involved in the algebraic expression of Cui and Cao's APA S-box are listed in the following table:

| | **0** | **1** | **2** | **3** | **4** | **5** | **6** | **7** | **8** | **9** | **A** | **B** | **C** | **D** | **E** | **F** |
|---|---|---|---|---|---|---|---|---|---|---|---|---|---|---|---|---|
| **0** | 8C | 3B | AA | 1A | F7 | F9 | 12 | 92 | 8A | A9 | 49 | 0E | DD | 25 | F9 | FB |
| **1** | FA | 2B | 85 | 27 | 75 | E5 | 64 | F7 | 36 | BB | 33 | D2 | 9C | 7D | C9 | B0 |
| **2** | 32 | 01 | E0 | 2E | 69 | 01 | AA | 3C | DB | 19 | 92 | 33 | 00 | 6C | 32 | 8D |
| **3** | 5B | A6 | B7 | D0 | 76 | 65 | C4 | EA | B5 | 81 | 74 | EC | 08 | A6 | 9F | 62 |
| **4** | 03 | 0F | DB | 98 | 64 | 75 | DC | 40 | 08 | DD | FD | EA | 0E | B4 | F3 | E2 |
| **5** | A5 | 48 | 23 | 75 | 5B | 7D | 17 | 4C | 4C | AE | 72 | 63 | DA | 49 | E8 | 9E |
| **6** | B0 | 5E | A9 | 7F | E3 | 2B | 14 | 57 | 51 | 05 | 00 | B3 | 51 | A2 | 99 | 31 |
| **7** | 3C | 36 | 44 | F1 | 8A | 05 | 62 | 50 | 68 | 83 | 41 | 65 | C9 | 3D | F0 | 4D |
| **8** | A0 | E1 | BE | 8E | 73 | D5 | A7 | C2 | BF | 3E | 91 | D4 | 79 | D6 | CA | D8 |
| **9** | 90 | D0 | A0 | 93 | D5 | 03 | E5 | 7F | 3F | 9F | 27 | 1D | 32 | 90 | CD | F3 |
| **A** | B7 | 2F | 1E | 0A | 5A | 5C | 20 | CF | 13 | AD | B1 | 84 | A4 | 98 | C8 | C3 |
| **B** | 13 | 3D | 96 | A6 | 77 | 92 | 6D | 64 | D0 | BE | 4A | 3A | 39 | FD | 78 | DF |
| **C** | CE | 49 | 88 | 75 | 81 | 4B | FE | 0D | A3 | 08 | 69 | 1B | B2 | 2A | 5E | 11 |
| **D** | 3D | 92 | CB | 68 | 0C | 6E | 1E | B3 | E5 | C2 | 05 | 1E | 2F | 7F | B1 | 05 |
| **E** | 14 | 77 | E7 | 78 | F5 | E2 | 3A | 94 | A6 | 66 | B1 | 5C | DD | 6F | 6B | D7 |
| **F** | B0 | E0 | 91 | EF | 3B | E9 | 80 | 57 | AA | 59 | 91 | 97 | 21 | 5C | 2F | 00 |

Clearly, the algebraic expression of the AES S-box contains only 9 terms while the affine-power-affine S-box contains 253 non-zero coefficients thereby making the S-box more resistant to algebraic attacks. The affine-power-affine S-box also inherits other cryptographic characteristics of AES S-box.

### 3.4 Keyed Latin Square

Latin square of order *n* refers to a *n×n* grid containing n distinct symbols such that every row and every column contains distinct entries. Such a square is termed as "Latin square" because *Leonhard Euler*, the person who developed it, used Latin symbols to design this square. The most common example of Latin square in our daily life is the game of Sudoku, which has an additional constraint on the block. A Latin square of size 5×5 is depicted in Fig. 1. For a given order n, there can be a huge number of possible Latin squares. For example, there can be approximately $10^{37}$ Latin squares for order 10. This expounds the use of Latin squares in image encryption algorithms. Wu et al. [9], in 2013, depicted the drawbacks of the chaotic-based image encryption techniques and came up with an image encryption algorithm based on Latin Square. They proposed the algorithm for developing a 256×256 keyed Latin Square, based on a 256 bit external key, which was then used to encrypt the image. The readers can refer to [9, 19] for a detailed description of Latin Square. We adopt the keyed Latin square generation algorithm presented by Wu et al. in our work. Thus, we combine the merits of both the chaotic systems as well as the Latin square to develop an effective image encryption algorithm.

| 2 | 0 | 4 | 1 | 3 |
|---|---|---|---|---|
| 3 | 2 | 1 | 4 | 0 |
| 1 | 4 | 0 | 3 | 2 |
| 4 | 3 | 2 | 0 | 1 |
| 0 | 1 | 3 | 2 | 4 |

**Figure 1.** A Latin square of size 5×5

### 3.5 Fisher-Yates Shuffle

Fisher-Yates shuffling algorithm was proposed by Ronald Fisher and Frank Yates in 1938 to provide an algorithm for shuffling a linear array of finite length randomly [20]. The cardinal characteristic of this shuffling algorithm is that it is completely evenhanded i.e. all the permutations of the array are equally likely. To make it computer implementable, Richard Durstenfeld modified the algorithm so as to lower its time complexity from $O(n^2)$ to $O(n)$ and hence this modified algorithm is referred to as the modern version of Fisher-Yates shuffle. The modern version is also space efficient requiring no additional storage space. The modern Fisher-Yates shuffling algorithm is depicted below:

To shuffle an array *S* of *n* elements (indices 1...*n*)

> *for i ← n downto 1 do*
>    *j ← random integer with $1 \leq j \leq i$*
>    *exchange(S[j], S[i])*
> *end*

The core part of the algorithm which makes it unbiased is the generation of the indices *j*. In our algorithm, we use Fisher-Yates shuffle to permute an initial array of 256 elements which are the elements of the 8×8 S-box. The shuffling technique is improved by incorporating piecewise linear chaotic map for random generation of indices.

## 4. PROPOSED ALGORITHM

### 4.1 Dynamic S-Boxes Generation

The proposed methodology for the design of dynamic S-boxes is as follows:

**S.1.** Initialize a linear array *S* of size 256 with values starting from 0 to 255 in ascending order. Set $\Delta = length(S)$.
**S.2.** Iterate the piece-wise linear chaotic map of Eqn. (2) for $N_0$ times to die-out the transient effect of map with selected initial conditions.
**S.3.** Set *cnt* = 1.
**S.4.** Further iterate the map (1) and sample the chaotic state-variable *x*.
**S.5.** Extract a random number $m \in [1, k]$ from *x* as:

$$m = \{floor(x*10^{10})\}mod(k) +1 \qquad where, \ k = \Delta - cnt+1$$

**S.6.** Exchange the two elements of array *S* at positions *m* and *k* i.e. $S(k) \leftrightarrow S(m)$.
**S.7.** Set *cnt* = *cnt* + 1, If *cnt* < 256 go to step **S.4**.
**S.8.** Re-apply the Fisher-Yates shuffle on current array *S* by repeating steps **S.3** to **S.7** for *ξ* times.
**S.9.** Translate resultant shuffled linear array *S* to 16×16 table to denote final S-box.

The key drives dynamic generation of S-box. A number of different S-boxes can be generated by making a minor change in the key. The above method is applied to construct 1000 different S-boxes by updating the initial value of *x(0)* for Eqn. (2) with an increment of 0.000223 for each S-box every time. Like other S-box construction methods, the proposed scheme is also simple having low algorithmic complexity and low computational cost. Moreover, our scheme generates efficient S-boxes having better cryptographic features as compared to many other existing schemes. The performance of all 1000 S-boxes has been experimentally investigated, statistically analyzed and reported by one of the author in [21].

### 4.2 Image Encryption Algorithm

In this algorithm, we encrypt the image by combining and exploring the virtues of chaos, substitution-boxes, affine-power-affine transformation and keyed Latin square. The S-Box and the element of S-Box are randomly selected using the digits of chaotic map variable. The pixels of the image are then confused by performing XOR operations on the pixel value, the output of affine-power-affine transformation, the previous cipher pixel value and the value of Latin square. When all pixels have been modified, the resultant image is rotated and flipped about its left diagonal. The complete procedure is repeated for a number of rounds to get encrypted image. The steps of proposed image encryption algorithm through chaotic substitution are as follows:

**A.1.** Generate 1000 dynamic S-boxes as discussed in Section 4.1.
**A.2.** Input plain-image *P* (having size M×N) the secret key components *β* (number of rounds), *x(0)*, *C(0)* and Latin square key *K*.
**A.3.** Let *r* be the count variable for round operation and initially *r* = 1.
**A.4.** Update the key components *x(0)*, *C(0)* and *K* using current count *r*.
**A.5.** Reshape the image into a 1D array *P* and set *i* = 1.
**A.6.** Generate 256×256 Latin square using key *K* and reshape it to 1D array *L*.
**A.7.** Iterate chaotic Logistic map of Eqn.(1) once and capture the map variable *x* of the form $x = 0.d_1d_2d_3 \ldots d_{15}$ ($0 \leq d_i \leq 9$). Calculate the following:

$$k = (a_1) mod(1000) + 1$$
$$l = (a_2) mod(16) + 1$$
$$m = (a_3) mod(16) + 1$$

where
$$a_1 = d_1d_2d_3d_4d_5$$
$$a_2 = d_6d_7d_8d_9d_{10}$$
$$a_3 = d_{11}d_{12}d_{13}d_{14}d_{15}$$

**A.8.** Take APA-transformation of value obtained from $k^{th}$ S-Box at the index (*l*, *m*). Let this value be *Φ*.
**A.9.** Calculate the cipher image pixel as

$$C(i) = C(i-1) \oplus P(i) \oplus \Phi \oplus L(q)$$

where $q = (i) mod(256 \times 256) + 1$

**A.10.** Evaluate $t = (C(i)) mod(4) + 1$ and iterate the chaotic Logistic map for *t* times and discard the values.
**A.11.** If *i* < M×N then set *i* = *i* + 1 and go to step **A.7**.
**A.12.** Reshape cipher image *C* into 2D matrix and perform two 90º anti-clockwise rotations and then flip about its left diagonal.
**A.13.** If *r* < *β* then set *r* = *r* +1, *P* = *C* and go to step **A.4** else terminate.

The block diagram of the proposed image encryption method is demonstrated in Figure 2. The decryption process is similar to the encryption process but all the operations are applied in reverse order. Firstly, generate dynamic S-boxes and then start with the last round first i.e. move in reverse direction. Before the start of each round, we flip the image matrix about its left diagonal and then rotate the matrix twice by 90º clockwise. Then we apply the operations, as in the encryption process, to all the pixels of the image and this is continued till all rounds are complete. Thus, at the end original image is obtained.

## 5. SIMULATION RESULTS

The proposed scheme is simulated on Matlab 2011a with three plain-images each of size 256×256, which are shown in Figure 3 and their histograms are shown in Figure 4. The initial value used for encrypting the images are: $x(0) = 0.23456$, $\lambda = 3.99$, $K =$ '12A34F56E78D90C31B72AF4835DC0981237654CD185A3FEB01CAE7259018FD14', $\beta = 4$. The corresponding encrypted images by the proposed algorithm are shown in Figure 5 and histograms are shown in Figure 6. It is evident that images in Figure 5 have high visual indistinguishability and distortion. The encrypted images, by the Anees *et al.* encryption method, with their corresponding histograms are depicted in Figure 7 and 8 respectively. The histogram of an image depicts how the pixels in the image are distributed. A perfectly encrypted or noise-image has a flat and uniform histogram. As can be seen that proposed algorithm provides better encryption effect than the Anees at al algorithm. Reason being the encrypted images shown in Figure 5 are more distorted visually than the ones shown in Figure 7. Moreover, the histograms of our encrypted images shown in Figure 8 are more uniform and flat akin to noise image than the histograms of images encrypted with Anees *et al.* algorithm.

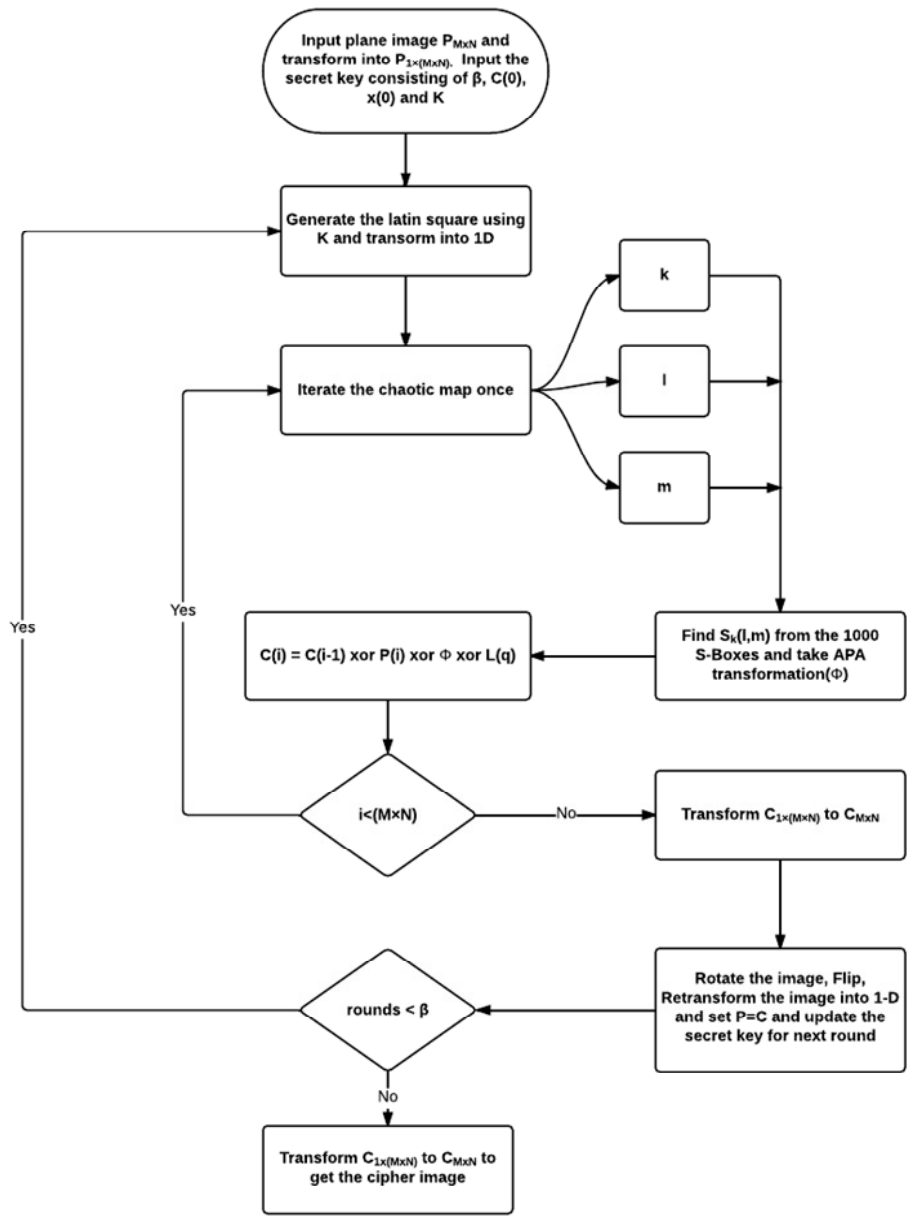

**Figure 2.** Flowchart depicting the proposed image encryption algorithm

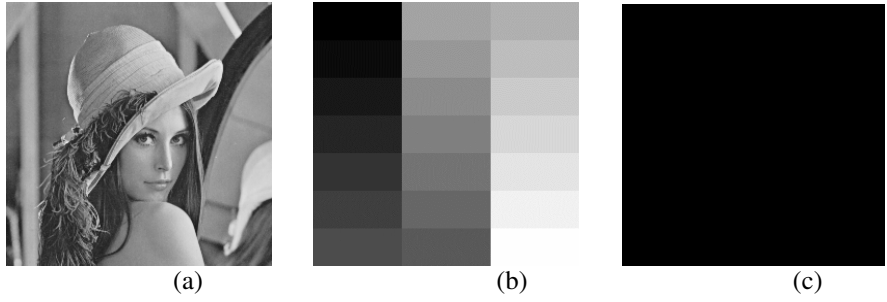

**Figure 3.** Plain-images: (a) *Lena* (b) *Gray-strips* (c) *Black*

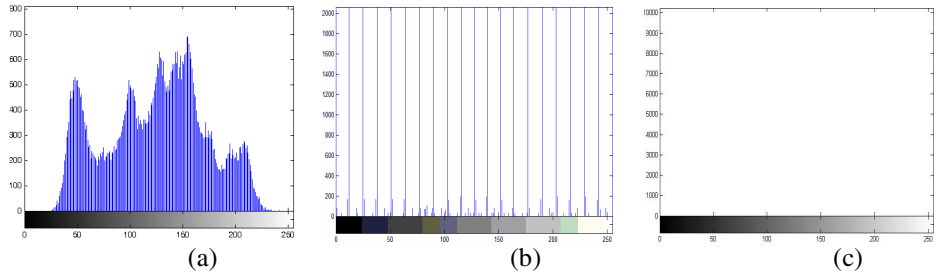

**Figure 4.** Histograms of plain-images: (a) *Lena* (b) *Gray-strips* (c) *Black*

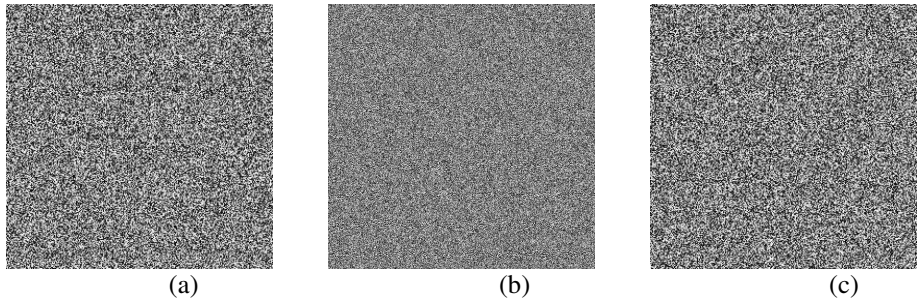

**Figure 5.** Encrypted images with proposed algorithm: (a) *Lena* (b) *Gray-strips* (c) *Black*

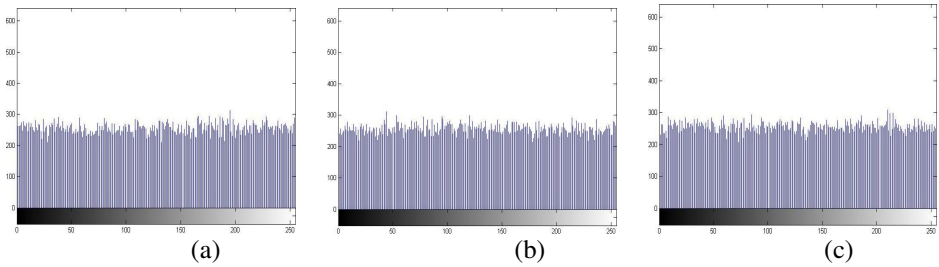

**Figure 6.** Histograms of encrypted images in Figure 5 : (a) *Lena* (b) *Gray-strips* (c) *Black*

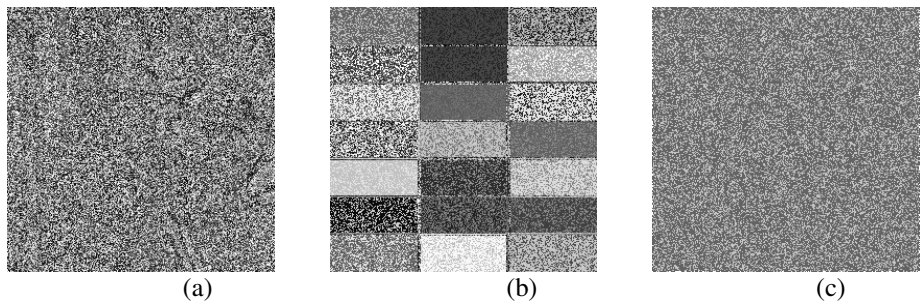

**Figure 7.** Encrypted images with Anees *et al.* algorithm: (a) *Lena* (b) *Gray-strips* (c) *Black*

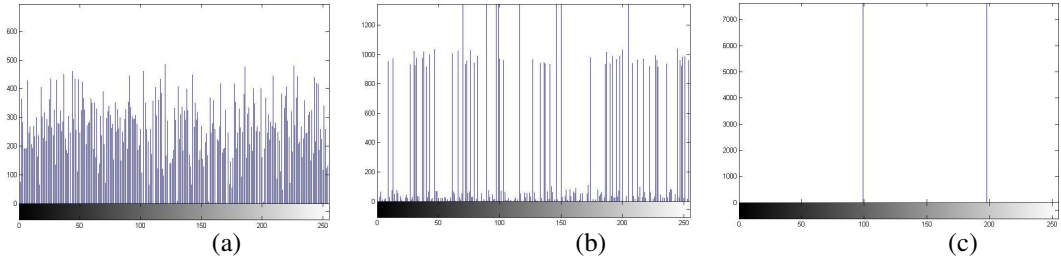

**Figure 8.** Histograms of encrypted images in Figure 7 : (a) *Lena* (b) *Gray-strips* (c) *Black*

### 5.1 Key Space

In proposed algorithm, the components of secret key includes *x(0), λ, β, C(0), K* and key used for dynamic S-boxes generation. With a 15 digit precision, the size of key space comes out as $6 \times 256 \times 1000 \times 10 \times 2^{256} \times (10^{14})^4 \approx 2^{465}$. Thus, the key space is large enough to resist the brute-force attack.

### 5.2 Pixels Auto-Correlation

The correlation analysis is done to measure the amount of correlation that is present among adjacent pixels in an image. When the value of the correlation test is close to 1, the images are positively linearly correlated. On the contrary, when the value of the correlation is close to 0, the images are not correlated to each other. Thus, for a strong encryption algorithm, the correlation between the encrypted image and the original image should be as close to 0 as possible [9, 22]. Correlation between the images is calculated using the formula:

$$\rho = \frac{N \sum_{i=1}^{N}(x_i \times y_i) - \sum_{i=1}^{N} x_i \times \sum_{i=1}^{N} y_i}{\sqrt{(N \sum_{i=1}^{N} x_i^2 - (\sum_{i=1}^{N} x_i)^2) \times (N \sum_{i=1}^{N} y_i^2 - (\sum_{i=1}^{N} y_i)^2)}} \qquad (3)$$

The correlation measures for images under consideration are listed in Table 1.

**Table 1.** Correlation coefficients in different images

| Image name | Plain-image | Ref. [4] | Proposed |
|---|---|---|---|
| Lena | 0.95679 | 0.00846 | 0.00673 |
| Gray-strips | 0.99979 | 0.3944 | -0.00128 |
| Black | 1 | 0.01712 | 0.00623 |

### 5.3 Image Entropy

Entropy analysis is done to measure the amount of randomness that can be used to describe the image texture and information content [22-26]. It refers to the image pixels ability to adopt various gray levels. The entropy is high if the image pixels can adopt many gray levels while the entropy is low if the image pixels can adopt only a few gray levels. The entropy is calculated using the formula:

$$H(S) = -\sum_{i=0}^{255} p(s_i) \log_2(p(s_i))$$

(4)

Where $p(s_i)$ represents the probability of symbol $s_i$ and the entropy is expressed in bits. If the source $S$ emits $2^8$ symbols with equal probability, i.e. $S = \{s_0, s_1, \ldots, s_{255}\}$ then the entropy is $H(S) = 8$, which corresponds to a true random source and represents ideal value of entropy for message source $S$. If the entropy of an encrypted image is significantly less than the ideal value 8, then, there would be a possibility of predictability which threatens the image security. The information entropy obtained for original and encrypted images are given in Table 2. The entropy measures obtained for encrypted images show that the information leakage by the proposed encryption scheme is negligible and is secure against the entropy attack.

**Table 2.** Entropy measures of images

| Image name | Plain-image | Ref. [4] | Proposed |
|---|---|---|---|
| Lena | 7.4439 | 7.8403 | 7.9965 |
| Gray-strips | 4.3923 | 6.1582 | 7.9958 |
| Black | 0.0000 | 0.9185 | 7.9972 |

**5.4 Net Pixels Change Rate**

Net pixels change rate is used to quantify the plaintext sensitivity i.e. effect of changing a single pixel in original image on encrypted image [9]. It also depicts the randomness and difference between the plain-image and its encrypted image. We take two encrypted images, $C_1$ and $C_2$, whose corresponding original images have only one-pixel difference. An array $D(i, j)$ is determined using $C_1$ and $C_2$. If $C_1(i, j) \neq C_2(i, j)$ then $D(i, j) = 1$, otherwise $D(i, j) = 0$. The NPCR is defined as:

$$NPCR = \frac{\sum_{i=1}^{M}\sum_{j=1}^{N} D(i, j)}{M \times N} \times 100 \tag{5}$$

Larger the value of NPCR better is the plain image sensitivity offered by the encryption algorithm. The NPCR measures verify that the proposed encryption algorithm is more sensitive to a slight change in the plain-image. Thus, the proposed algorithm exhibits better resistance against the differential attack.

**Table 3.** NPCR measures of images

| Image name | Ref. [4] | Proposed |
|---|---|---|
| Lena | 0.00107 | 34.37 |
| Gray-strips | 0.00121 | 36.72 |
| Black | 0.00153 | 99.60 |

**6. CONCLUSION**

In this paper, we presented a new image encryption algorithm which uses Fisher-Yates shuffle and chaos to first generate 1000 dynamic S-boxes and then uses these S-boxes along with Latin square, chaos and affine-power-affine structure to generate encrypted image. Our algorithm provides high security as it uses chaotic Logistic map to select randomly an S-box for substitution of image pixel which is substituted with Latin square, generated from an external key, to produce pixel of encrypted image. More than two rounds of encryption algorithm are recommended to ensure that the cipher image has no correlation with the original image and increased robustness of algorithm. The chaotic map also makes encryption process key dependent. The key space of the algorithm is extremely high which clearly suggests that our algorithm is resistant to brute-force attack. Moreover, the experimental results and various security analyses suggest that proposed algorithm is resistant to statistical attacks, known-plaintext attack and chosen plaintext attack. Proposed algorithm outperforms when compared with recent image encryption technique which confirms its excellent practicableness and suitableness.